\documentstyle[12pt,aaspp4,flushrt]{article}

\slugcomment{Submitted to The Astrophysical Journal}

\catcode`\@=11 
\def\@versim#1#2{\vcenter{\offinterlineskip
        \ialign{$\m@th#1\hfil##\hfil$\crcr#2\crcr\sim\crcr } }}
\newcommand{\beq}{\begin{equation}}
\newcommand{\eeq}{\end{equation}}
\def\lsim{\mathrel{\mathpalette\@versim<}}
\def\gsim{\mathrel{\mathpalette\@versim>}}

\def\om{\tilde \omega}

\begin{document}
\title{The Magnetohydrodynamics of Convection-Dominated Accretion
Flows} 
\author{Ramesh Narayan\altaffilmark{1}} 

\affil{Department of Astrophysical Sciences, Princeton University,
Princeton, NJ 08540; narayan@cfa.harvard.edu}

\altaffiltext{1}{Permanent Address: Department of Astronomy,
Harvard-Smithsonian Center for Astrophysics, 60 Garden Street,
Cambridge, MA 02138}

\author{Eliot Quataert}

\affil{Department of Astronomy, University of California, 501 Campbell
Hall, Berkeley, CA 94720; eliot@astron.berkeley.edu}

\author{Igor V. Igumenshchev}

\affil{Laboratory for Laser Energetics, University of Rochester, 250
East River Road, Rochester, NY 14623; iigu@lle.rochester.edu}

\author{Marek A. Abramowicz}

\affil{Department of Astronomy and Astrophysics, G\"oteborg
University/Chalmers University of Technology, S-41296, G\"oteborg,
Sweden; marek@fy.chalmers.se}

\medskip

\begin{abstract}

Radiatively inefficient accretion flows onto black holes are
unstable due to both an outwardly decreasing entropy (``convection'')
and an outwardly decreasing rotation rate (the ``magnetorotational
instability''; MRI).  Using a linear magnetohydrodynamic
stability analysis, we show that long-wavelength modes with $\lambda/H
\gg \beta^{-1/2}$ are primarily destabilized by the entropy gradient
and that such ``convective'' modes transport angular momentum {\it
inwards} ($\lambda$ is the wavelength of the mode, $H$ is the scale
height of the disk, and $\beta$ is the ratio of the gas pressure to
the magnetic pressure).  Moreover, the stability criteria for the
convective modes are the standard H{\o}iland criteria of hydrodynamics.
By contrast, shorter wavelength modes with $\lambda/H \sim
\beta^{-1/2}$ are primarily destabilized by magnetic tension and
differential rotation.  These ``MRI'' modes transport angular momentum
{\it outwards.}  The convection-dominated accretion flow 
(CDAF) model, which has been proposed for radiatively inefficient
accretion onto a black hole, posits that inward angular momentum
transport and outward energy transport by long-wavelength convective
fluctuations are crucial for determining the structure of the
accretion flow. Our analysis suggests that 
the CDAF model is applicable to a magnetohydrodynamic accretion flow
provided the magnetic field saturates at a sufficiently
sub-equipartition value ($\beta \gg 1$), so that long-wavelength
convective fluctuations with $\lambda/H \gg \beta^{-1/2}$ can fit
inside the accretion disk.  Numerical magnetohydrodynamic simulations
are required to determine whether such a sub-equipartition field is in
fact obtained.

\

\noindent {\it Subject Headings:} accretion --- accretion disks ---
black hole physics --- convection --- instabilities ---MHD ---
turbulence

\end{abstract}

\section{Introduction}

Models of radiatively inefficient accretion flows provide a useful
framework for interpreting observations of low-luminosity black hole
X-ray binaries and active galactic nuclei (see, e.g., Narayan,
Mahadevan \& Quataert 1998; Quataert 2001; Narayan 2002 for reviews).
In the past few years, there has been rapid theoretical progress in
understanding the dynamics of such flows.  Much of this advance has
been driven by numerical simulations.  In particular, a number of
hydrodynamic simulations have been reported in which angular momentum
transport is put in ``by hand'' using an $\alpha$ prescription (e.g.,
two-dimensional simulations by Igumenshchev, Chen, \& Abramowicz 1996;
Igumenshchev \& Abramowicz 1999, 2000; Stone, Pringle, \& Begelman
1999; and three-dimensional simulations by Igumenshchev, Abramowicz \&
Narayan 2000).  Such simulations reveal a flow structure very
different from advection-dominated accretion flow models (ADAFs) that
were proposed to describe the structure of radiatively inefficient
accretion flows (Ichimaru 1977; Narayan \& Yi 1994; Abramowicz et
al. 1995).  In an ADAF, the gas accretes rapidly, and the
gravitational potential energy of the accreting gas is stored as
thermal energy and advected into the central black hole.  By contrast,
in the hydrodynamic simulations, the rate of mass accretion is much
smaller and strong radial convection efficiently transports energy
outwards.  The simulations have been interpreted in terms of a
``convection-dominated accretion flow'' model (CDAF; Narayan,
Igumenshchev \& Abramowicz 2000; Quataert \& Gruzinov 2000; Abramowicz
et al. 2002).  In a CDAF, convection simultaneously transports energy
outwards and angular momentum inwards, strongly suppressing the
accretion rate onto the central black hole.

The relevance of the hydrodynamic simulations, and the CDAF model
derived from them, is unclear.  This is primarily because of the {\it
ad hoc} treatment of angular momentum transport.  It is believed that
angular momentum transport in accretion flows is primarily due to MHD
turbulence initiated by the magnetorotational instability (MRI; Balbus
\& Hawley 1991; BH91).  Balbus \& Hawley have argued that, because of
the fundamental role played by magnetic fields, hydrodynamic models
typically cannot describe the structure of the accretion flow (or
differentially rotating systems more generally; e.g., Balbus \& Hawley
1998; Balbus 2000; 2001).  They have applied this criticism in detail
to the CDAF model (Hawley, Balbus, \& Stone 2001; Balbus \& Hawley
2002; hereafter BH02).  
In an independent argument, BH02 also suggest
that CDAF models violate the second law of thermodynamics.  We do not
consider this issue here, but intend to deal with it in a separate
paper.

In this paper we discuss the physics of the CDAF model within the
framework of MHD.  Apart from clarifying the theoretical underpinnings
of the CDAF model, we believe that our analysis also provides a useful
illustration of the conditions under which a hydrodynamic model is
applicable to an intrinsically magnetohydrodynamic situation; this is
important in other astrophysical contexts, e.g., the excitation of
density waves at Lindblad resonances in a magnetized accretion flow,
or diskoseismology models of QPOs.

The paper is organized as follows.  In \S2 we discuss the linear
stability of a differentially rotating and thermally stratified plasma
in the presence of a weak vertical magnetic field.  We identify and
explain the difference between ``convective'' modes and ``MRI'' modes
and show that the former transport angular momentum inwards while the
latter usually transport angular momentum outwards.  Then, in \S2.1 we
generalize this result to an arbitrary field orientation and arbitrary
stratification.  In \S3 we discuss the implications of our results.
In particular, we use the linear analysis to speculate about the
conditions under which the saturated nonlinear turbulence in an
accretion flow might give rise to a CDAF-like structure. In \S4 we
summarize our results.

\section{Convective Modes in a Magnetized Differentially Rotating Plasma}

We consider the linear stability of a differentially rotating and
thermally stratified plasma in the presence of a weak magnetic field.
We specialize to the case of a vertical magnetic field, $B_z \equiv
B$, and modes with purely vertical wavevectors, $k_z \equiv k$.  We
also take the rotation rate, pressure, and density to be constant on
cylinders, i.e., $\Omega(R)$, $P(R)$, and $\rho(R)$.  We thus employ
the same equations as those in BH02 (which are a simplification of the
more general analysis in BH91).  Although the problem we analyze is
identical to that considered by BH02, we differ 
in the
interpretation of the results.  In particular, we identify a set of
modes that have all the features of growing convective modes in
hydrodynamics, including inward transport of angular momentum. 


The thermal stratification of the flow can be described by the
Brunt-V{\"a}is{\"a}l{\"a} frequency, \beq
N^2=-N_R^2={3\over5\rho}{\partial P\over\partial R} {\partial\ln
P\rho^{-5/3}\over\partial R}. \label{brunt} \eeq In the absence of
rotation, the system is convectively unstable if $N^2 > 0$.  In a
rotating medium, but ignoring magnetic fields, convection is present
only if $N^2 > \kappa^2$, where $\kappa^2 = 4 \Omega^2 + d \Omega^2 /d
\ln R$ is the epicyclic frequency.\footnote{The requirements for
convection in an unmagnetized rotating medium are the H{\o}iland
criteria (e.g., Tassoul 1978).  These reduce to the condition $N^2 >
\kappa^2$ for the case considered here.  For simplicity, we will refer
to this condition as the H{\o}iland criterion in this section.  \S2.1
treats the general case.}  Note that $N^2 > \kappa^2$ requires a
sound speed comparable to the rotation speed.

For a weakly magnetized medium, BH91 derived the behavior of linear
perturbations including the effects of thermal stratification.  For
fluid displacements of the form ${\bf \xi} \propto \exp[ikz + \gamma
t]$, and for the particular geometry considered here, the growth rate
$\gamma$ is given by \beq \gamma^2 = -(k v_A)^2 + {1\over2}\left[N^2 -
\kappa^2 \pm \sqrt{(N^2-\kappa^2)^2 + 16\Omega^2(kv_A)^2}\right],
\label{growth} \eeq where $v_A = B/\sqrt{4 \pi \rho}$ is the Alfv\'en
speed.  For the remainder of this section we focus on the growing mode
in equation (\ref{growth}); this has $\gamma^2 > 0$ and corresponds to
the positive sign of the square root term.  In the Appendix we present
a more general analysis of both growing and oscillatory modes that
helps clarify some of the issues discussed in this section.

To understand the physics behind the dispersion relation in equation
(\ref{growth}) consider the equations for the radial and azimuthal
displacements of fluid elements (BH91): \beq {\partial^2 \xi_R \over
\partial t^2} - 2 \Omega {\partial \xi_\phi \over \partial t } =
-\left({d \Omega^2 \over d \ln R} + (kv_A)^2\right)\xi_R + N^2 \xi_R ,
\label{radial} \eeq \beq {\partial^2 \xi_\phi \over \partial t^2} + 2 
\Omega {\partial \xi_R \over \partial t } = -(kv_A)^2
\xi_\phi. \label{azimuthal} \eeq Solving equation (\ref{azimuthal})
for $\xi_\phi$ we can substitute into equation (\ref{radial}) to find
\beq {\partial^2 \xi_R \over \partial t^2} = a_H + a_{M,s} + a_{M,d},
\label{radialtot} \eeq 
where the three acceleration terms are of the form \beq a_H=(N^2 -
\kappa^2)\xi_R, \qquad a_{M,s} = - (kv_A)^2 \xi_R, \qquad a_{M,d} =
\left[{4 \Omega^2 k^2 v_A^2 \over \gamma^2 +
k^2v_A^2}\right]\xi_R. \label{terms} \eeq Since $\partial^2
\xi_R/\partial^2 t = \gamma^2 \xi_R$ it is straightforward to solve
equation (\ref{radialtot}) and obtain the dispersion relation in
equation (\ref{growth}).

The first acceleration term in equation (\ref{radialtot}), $a_H$, is a
pure ``hydrodynamic'' or ``H{\o}iland'' term (hence the subscript $H$)
which is present even in the absence of magnetic fields.  This term
can be either stabilizing or destabilizing; it is stabilizing when
$N^2<\kappa^2$ (the medium is convectively stable by the H{\o}iland
criteria) and destabilizing when $N^2>\kappa^2$ (convectively
unstable).  The second and third terms in equation (\ref{terms}) owe
their existence to the magnetic field (hence subscript $M$).  One of
them is always stabilizing (subscript $s$) and the other is always
destabilizing (subscript $d$).

Equations (\ref{radial})--(\ref{terms}) have a very clean physical
interpretation, as we explain in the rest of this section.  For the
simple geometry considered here (vertical field and vertical
wavevector), gas and magnetic pressure forces are unimportant.  If
$N^2 = 0$, the only forces in the problem are magnetic tension and
rotating frame dynamics (e.g. the Coriolis force).  Equations
(\ref{radial}) and (\ref{azimuthal}) then describe the behavior of
fluid elements that are coupled by the tension of magnetic field
lines.  As Balbus \& Hawley (1992, 1998) have discussed, these
equations are identical to those describing fluid elements coupled by
a spring of spring constant $(kv_A)^2$.  For a stratified medium with
$N^2 \ne 0$, the radial acceleration of a fluid element has an
additional component due to the radial buoyancy force.  This is the
$N^2 \xi_R$ term in equations (\ref{radial}) and (\ref{radialtot}).
When $N^2<\kappa^2$, this term does not have a large effect and
introduces modest changes to the values of numerical factors.
However, when $N^2>\kappa^2$, which corresponds to a convectively
unstable system by the H{\o}iland criterion, the acceleration term
$a_H$ becomes positive and destabilizing.  This introduces
qualitatively new effects.

The angular momentum flux of a linear mode is given by $R \Omega
T_{R\phi}$, where $T_{R \phi}$ is the $R-\phi$ component of the fluid
stress tensor, \beq T_{R \phi} = \rho(\delta v_R \delta v_\phi -
\delta v_{AR} \delta v_{A\phi}), \eeq $\delta {\bf v}$ is the
perturbed fluid velocity, and $\delta {\bf v_A} \equiv \delta {\bf
B}/\sqrt{4 \pi \rho}$.  Using the eigenfunctions in BH91 we confirm
BH02's expression for the stress tensor, namely \beq {T_{R\phi} \over
\rho |\delta v_R^2|} = {\Omega \over D \gamma} \left[{k^2v_A^2 \over
\gamma^2}\left(\left|{d\ln \Omega \over d \ln R}\right| + 2 \right)-
{\kappa^2 \over 2 \Omega^2}\right] \equiv \left({\Omega \over \gamma}
\right) t_{R \phi}, \label{stress} \eeq where $D = 1 +
k^2v_A^2/\gamma^2$ and $t_{R\phi}$ is a useful dimensionless stress
whose sign gives the sign of $T_{R \phi}$.  

\begin{figure}
\plotone{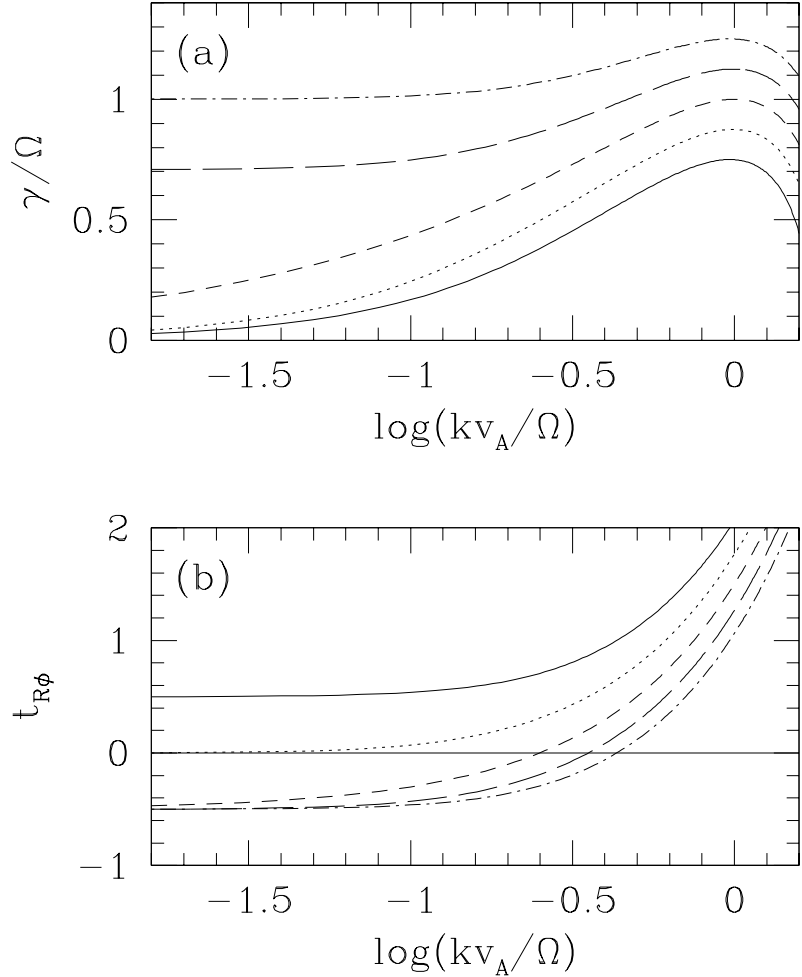}
\caption{(a) Dimensionless growth rate $\gamma/\Omega$
as a function of dimensionless wavevector $kv_A/\Omega$ in a Keplerian
system ($\kappa^2=\Omega^2$), for five choices of the Brunt-V{\"a}is{\"a}l{\"a}
frequency: $N^2/\Omega^2 =$ 0 (solid line), 0.5 (dotted line), 1
(short-dashed line), 1.5 (long-dashed line), 2 (dot-dashed line).  The
system is unstable by the H{\o}iland criterion for $N^2/\Omega^2=$ 1.5
and 2.  For these two cases, the growth rate remains finite in the
long-wavelength limit $kv_A/\Omega\ll1$ and the modes are
indistinguishable from the corresponding unstable hydrodynamical
modes.  (b) Dimensionless shear stress corresponding to the same modes
as in (a).  For $N^2/\Omega^2=$ 1.5, 2, and $kv_A/\Omega\ll1$, the
stress is negative, implying that these convective modes transfer
angular momentum inwards, just as in hydrodynamics.
\label{fig1}}
\end{figure}

\placefigure{fig1}

We are now in a position to elucidate the physics of the linear
instability calculation.  Figure 1a shows a plot of the growth rate
$\gamma/\Omega$ as a function of the dimensionless wavevector
$kv_A/\Omega$ for 5 values of the Brunt-V{\"a}is{\"a}l{\"a} frequency:
$N^2/\Omega^2 = 0, 0.5, 1, 1.5,$ and $2$.  We have assumed a Keplerian
rotation profile, for which $\kappa^2 = \Omega^2$.  Thus, the first
two choices of $N^2$ correspond to a stable entropy gradient by the
H{\o}iland criterion, the third is neutrally stable, and the last two
choices correspond to an unstable entropy gradient.  Figure 1b shows
the dimensionless stress $t_{R \phi}$ as a function of $kv_A/\Omega$
for the same five values of $N^2$.  It is useful to note that the
self-similar ADAF model gives $N^2/\Omega^2 = 15(\gamma_{ad} -
1)/4\gamma_{ad}$, where $\gamma_{ad}$ is the adiabatic index of the
gas (Narayan \& Yi 1994).\footnote{This expression is not valid if
$\gamma_{ad} = 5/3$ since then $\Omega^2 = 0$ in the analytical
models.}  For example, if $\gamma_{ad} = 1.5$, $N^2/\Omega^2 = 1.25$.

For $N^2 = 0$, the results reproduce the usual MRI.  The growth rate
is very small for $k v_A \ll \Omega$ and peaks when $kv_A$ is $\approx
\Omega$, with a peak value of $\gamma = 3 \Omega/4$; the angular
momentum flux is always outwards ($T_{R\phi} > 0$).  The instability
is clearly triggered by the magnetic field via the destabilizing term
$a_{M,d}$, since the other two terms in equation (\ref{radialtot}) are
both stabilizing (negative).  The MRI survives with qualitatively the
same behavior even for non-zero $N^2$, so long as the H{\o}iland
criterion for convective stability ($N^2<\kappa^2$) is satisfied.

Consider, however, a flow with $N^2 > \kappa^2$.  Such a flow is
unstable to convection according to hydrodynamics, and is therefore
the case of interest for CDAFs.  Figure 1 shows that, for $N^2 >
\kappa^2$, modes with $k v_A \sim \Omega$ are not very different from
their $N^2 = 0$ counterparts.  The angular momentum flux is still
outwards and the growth rate of the mode is comparable (though
somewhat larger).  By contrast, long-wavelength modes are very
different.  As Figure 1 shows, long-wavelength modes are strongly
unstable and they transport angular momentum {\it inwards}.  In fact,
as can be seen by inspection, the $k v_A \ll \Omega$ limit of the
dispersion relation in equation (\ref{growth}) (with the + root) gives
$\gamma^2 = N^2 - \kappa^2$, and the corresponding stress tensor is
$T_{R \phi}/(\rho |\delta v_R|^2) = - \kappa^2/[2 \Omega (N^2 -
\kappa^2)^{1/2}] < 0$.  The long-wavelength modes are clearly
independent of the magnetic field and behave very differently from MRI
modes.

The physics of the $k v_A \ll \Omega$ modes is as follows.  In the
pure MRI problem, namely $N^2=0$ (or more generally $N^2<\kappa^2$),
long-wavelength modes are only weakly unstable because the magnetic
tension forces that are central to the MRI are weak; the field lines
are hardly bent by a long-wavelength perturbation.  By contrast, when
$N^2>\kappa^2$, the fluid has two destabilizing forces, the buoyancy
term ($a_H$) due to an unstable entropy gradient and the standard MRI
term ($a_{M,d}$).  The buoyancy force that drives convection is
independent of the wavelength of the perturbation, whereas the MRI
depends on $k$.  For long wavelengths, buoyancy is much more important
and controls all the physics of the mode, both the growth rate and the
angular momentum transport.  Moreover, we show in the Appendix that
the ``convective'' mode driven by buoyancy is the {\it only} unstable
long-wavelength mode in MHD for a H{\o}iland-unstable medium ($N^2 >
\kappa^2$).  In particular, long-wavelength perturbations for which
magnetic tension dominates (referred to as $-$ modes in the Appendix)
are stable oscillatory waves rather than unstable modes (see
eq. \ref{minus} and Fig. 3). 

\begin{figure}
\plotone{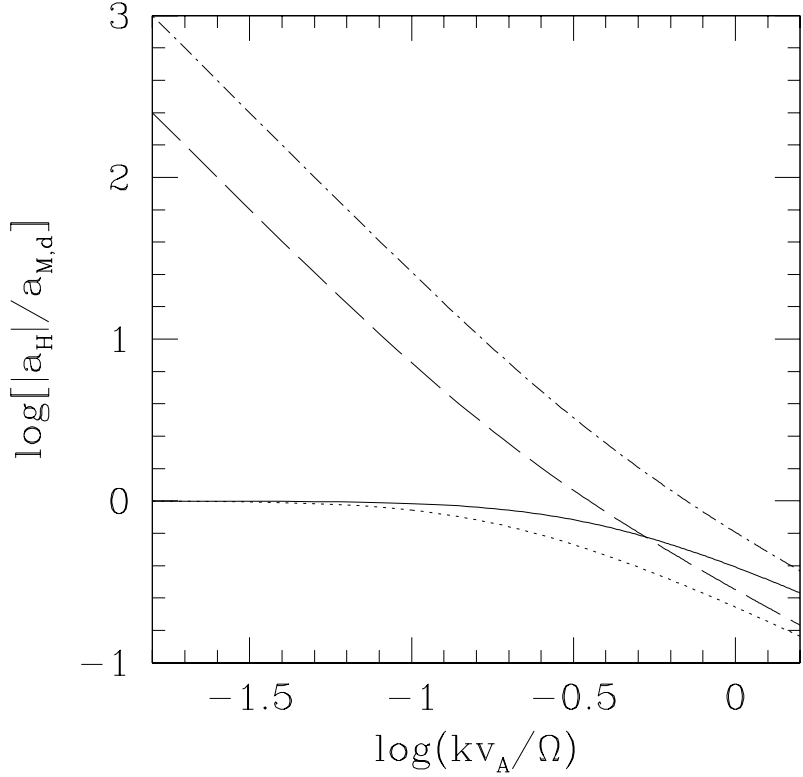}
\caption{Variation of $a_H/a_{M,d}$ versus
dimensionless wavevector $kv_A/\Omega$ for $N^2/\Omega^2=$ 0 (solid
line), 0.5 (dotted line), 1.5 (long-dashed line), 2 (dot-dashed line).
The ratio $a_H/a_{M,d}$ measures the importance of the hydrodynamic
buoyancy force relative to the leading destabilizing force due to
magnetic fields (see eqs. [\ref{radialtot}], [\ref{terms}]).  For the two
cases in which the system is convectively unstable according to the
H{\o}iland criterion, $viz., N^2/\Omega^2=$ 1.5, 2, the buoyancy force
clearly dominates at long wavelengths, confirming that these modes are
primarily convective, not magnetorotational.
\label{fig2}}
\end{figure}

\placefigure{fig2}

Figure 2 explicitly compares the relative importance of the two
destabilizing terms, the buoyancy term $a_H$ and the magnetic tension
term $a_{M,d}$.  Let us identify a mode as an ``MRI'' mode if
$a_H\lsim a_{M,d}$ and as a ``convective'' mode if $a_H>a_{M,d}$.
With this natural identification, we see that, for $N^2<\kappa^2$, all
unstable modes are MRI modes, regardless of the wavelength.  Even when
$N^2>\kappa^2$, short-wavelength modes with $kv_A\gsim\Omega$ are
still MRI modes.  However, unstable long-wavelength modes with
$kv_A\ll\Omega$ are clearly convective modes.  The growth of these
modes is due entirely to the unstable buoyancy force; correspondingly,
the modes transport angular momentum inwards (Fig. 1b), just as
indicated by hydrodynamic studies (e.g., Ryu \& Goodman 1992).  Indeed
these modes are virtually indistinguishable from their hydrodynamic
counterparts.  {\it All of the properties of {hydrodynamic} convective
instabilities therefore carry over to a {magnetohydrodynamic}
analysis.}
	

\subsection{The Hydrodynamic Limit}

Magnetic tension should be weak for fluctuations with $k v_A \ll
\Omega$ regardless of the details of the magnetic field geometry and
thermal stratification.  In this subsection, we show this explicitly
by considering the long-wavelength limit of the most general
axisymmetric MHD dispersion relation.

Balbus (1995) considered the linear, adiabatic, and axisymmetric
stability of a differentially rotating weakly magnetized plasma in the
presence of both vertical and radial stratification.  He showed that
the dispersion relation is given by \beq \om^4{k^2 \over k_z^2} -
\om^2\left[{3 \over 5 \rho} (D P)(D \ln [P \rho^{-5/3}]) + {1 \over
R^3} D(R^4\Omega^2)\right] - 4 \Omega^2 ({\bf k \cdot v_A})^2 = 0,
\label{disp} \eeq where \beq D = \left({k_R \over k_z}{\partial \over
\partial z} - {\partial \over \partial R}\right), \eeq and $\om^2 =
-\gamma^2 - ({\bf k} \cdot {\bf v_A})^2$ (for fluctuations $\propto
\exp[\gamma t + i {\bf k \cdot r}]$). Unlike in the previous section,
$\Omega$, $P$, and $\rho$ are now allowed to be functions of both $z$
and $R$, the magnetic field has an arbitrary direction (i.e., $B_R$,
$B_\phi$, and $B_z$ components), and the wavevector is given by ${\bf
k} = k_R {\bf \hat R} + k_z {\bf \hat z}$.

For ${\bf k \cdot v_A} = 0$ (but $k/k_z$, $k_R/k_z$, etc., finite),
equation (\ref{disp}) reduces to \beq \gamma^2 {k^2 \over k_z^2} +
\left[{3 \over 5 \rho} (D P)(D \ln [P \rho^{-5/3}]) + {1 \over
R^3}D(R^4\Omega^2)\right] = 0. \label{dispk0} \eeq Equation
(\ref{dispk0}) is independent of the magnetic field.  Indeed, as noted
by Balbus (1995), it describes the linear, adiabatic, axisymmetric,
hydrodynamic stability of a differentially rotating and thermally
stratified medium (e.g., Goldreich \& Shubert 1967).  As is readily
confirmed, equation (\ref{dispk0}) has unstable solutions if and only
if the H{\o}iland criteria are satisfied (e.g., Tassoul 1978).

The hydrodynamic limit in equation (\ref{dispk0}) formally requires
setting ${\bf k \cdot v_A} = 0$ in the MHD dispersion relation (eq.
[\ref{disp}]).  If ${\bf k \cdot v_A} \ne 0$, the full dispersion
relation (eq. [\ref{disp}]), which has twice the number of modes
(because it is fourth order in $\gamma$ rather than second order),
must be used.  However, if the medium is unstable by the H{\o}iland
criteria, and if we focus on long-wavelength MHD instabilities with
${\bf k \cdot v_A} \ll \Omega $, then the MHD system will behave like
its hydrodynamic counterpart; in particular, the only unstable mode is
effectively hydrodynamical and its growth rate is given by equation
(\ref{dispk0}) with small corrections $\sim ({\bf k \cdot
v_A}/\Omega)^2 \ll 1$.  Note that equation (\ref{dispk0}) describes
the unstable growing mode only when the medium is H{\o}iland unstable.
If the medium is H{\o}iland stable, the only possible instabilities
are intrinsically MHD in nature and equation (\ref{disp}) must be used
even for ${\bf k \cdot v_A} \ll \Omega$.  This point is explained in
more detail in the Appendix.

The analysis in this section shows that all of the properties of
axisymmetric hydrodynamic instabilities carry over to MHD in the
long-wavelength limit.  In particular, the results of the previous
section are general, and are not an artifact of the simplifying
assumptions made there.  Although our quantitative analysis is
restricted to axisymmetry, we suspect that long-wavelength
non-axisymmetric modes will behave hydrodynamically as well (since
they also do not significantly bend the magnetic field lines).

\section{Implications for CDAFs}

To assess the implications of our linear analysis it is useful to
consider wavelengths relative to the scale height of the disk, $H
\approx c_s/\Omega$, where $c_s$ is the sound speed.  The maximally
growing MRI mode has $k H \sim \beta^{1/2}$, where $\beta \approx
c^2_s/v^2_A$ is the ratio of the thermal energy to the magnetic energy
in the disk.  By contrast, the longer wavelength convective modes have
$k H \ll \beta^{1/2}$; more quantitatively, the linear analysis
suggests that buoyancy forces become important for $k H \lsim
0.3\beta^{1/2}$ (Fig. 2).  For these modes to be of interest they must
fit in the disk and so must have $k H \gsim 1$.  This in turn requires
$\beta \gg 1$, i.e., that the magnetic field must be sub-equipartition
($\beta \gsim 10$ may suffice).

Given the above analysis, we propose the following identification
between the hydrodynamic CDAF model and the present MHD results.  If
the magnetic field saturates at a sub-equipartition value, the most
unstable MRI mode operates on a scale $\sim \beta ^{-1/2} H \ll H$ and
transports angular momentum outwards. Moreover, if the medium is
H{\o}iland unstable, {\it only} such short-wavelength fluctuations
transport angular momentum outwards.  Longer wavelength convective
fluctuations are dominated by scales $\sim H$ and transport angular
momentum inwards (Fig. 1).

Which fluctuations are more important?  Or, more precisely, which part
of the power spectrum of fluctuations is more important, those with $k
H \sim 1$ or those with $k H \sim \beta^{-1/2}$?  This cannot really
be addressed analytically because we do not understand the saturation
of the linear instabilities (and it is clear that they are coupled;
e.g., convection, by stirring the fluid, can help to build up the
magnetic field).  Nonetheless, it is typically the case that the
energy in a turbulent plasma is dominated by the largest unstable
scales in the medium.  This suggests that, if $\beta \gg 1$ (perhaps
$\beta \gsim 10$ is sufficient), the larger-scale convective
fluctuations will dominate the dynamics, as proposed in the
hydrodynamic models.  Moreover, the inward angular momentum transport
by convective modes depends on the magnitude of $N^2 - \kappa^2$
(Fig. 1b).  Thus there is a natural way for the inward convective
transport to self-adjust to counteract outward transport by small
scale MRI fluctuations.  In the limit of efficient convection and
$\beta \gg 1$ we would expect marginal stability to convection
according to the H{\o}iland criteria, just as in the hydrodynamic models
(Quataert \& Gruzinov 2000).

It is important to stress several points: 

(1) Although we believe that our linear stability analysis sheds
important light on the physics of radiatively inefficient accretion
flows, it does not directly address the fully developed nonlinear
turbulence that is of primary importance to the accretion flow
structure.  This requires numerical simulations.

(2) The picture proposed here is very similar to the hydrodynamic
models of CDAFs; the primary change is that we have theoretical
support from a magnetohydrodynamic analysis.

(3) Balbus (2001) has shown that magnetized dilute plasmas can be
convectively unstable in the presence of an outwardly decreasing
temperature, rather than just an outwardly decreasing entropy.  It
would be interesting to incorporate this into future work.

(4) The extent to which one can usefully distinguish scales in the
power spectrum that are convective from those that are
magnetorotational depends on $\beta$, and becomes more and more useful
if $\beta \gg 1$.  It would be straightforward if the convective and
MRI modes occupied distinct and well-separated regions of wavevector
space.  Then even the nonlinear physics of the instabilities could be
distinct.  To cite an example from plasma physics, heat transport in
fusion devices is due to instabilities on both proton and electron
Larmor radii scales and one can often distinguish the physics of each
mechanism separately even in the nonlinear regime (e.g., Dorland et
al. 2000; Rogers, Dorland \& Kotschenreuther 2000).  As Figure 1
shows, such a separation in wavevector space is not present for our
problem.  On the other hand, Figs. 1 and 2 do show quite clearly that
the physics of the convective modes is very different from that of the
MRI modes.

(5) For a H{\o}iland-unstable medium ($N^2>\kappa^2$), there is a nice
continuity between the hydrodynamic limit and the strong field MHD
limit.  In wavevector space, the field-affected modes (the MRI modes)
are restricted to larger values of $k$ where magnetic tension is
important.  For very weak fields, these modes are at very large $k$,
and most of the long-wavelength modes that probably dominate the
dynamics are convective.  As the field strength increases, however,
the MRI modes expand downwards in $k$ space and become more and more
important.  Another way to look at this is as follows.  When the fluid
is unstable according to the H{\o}iland criterion, the switch that
determines whether or not convection is important is not the presence
or absence of a magnetic field, but rather the magnitude of the
dimensionless parameter $kv_A/\Omega\approx k H/\beta^{1/2}$.  When
$kv_A/\Omega$ is small, convection dominates the physics, whereas when
it is $\gsim1$, the magnetic field dominates.

\section{Conclusion}

In this paper we have used linear stability theory to assess the
conditions under which a hydrodynamic analysis is appropriate for an
intrinsically magnetohydrodynamic accretion flow.  We have shown that,
when the medium is unstable by the H{\o}iland criteria,
long-wavelength instabilities with $k v_A \ll \Omega$ are effectively
hydrodynamical.  Physically, this is because magnetic tension is
unimportant for sufficiently long-wavelength perturbations.  Formally,
the dispersion relation for a magnetohydrodynamic plasma reduces in
this limit to that of hydrodynamics (\S2). This shows that all of the
properties of hydrodynamic instabilities carry over to an MHD analysis
in the long-wavelength limit.

For the particular case of radiatively inefficient accretion flows
that are unstable to both convection and the MRI, the relevant
parameter that distinguishes convective behavior from
magnetorotational behavior is the quantity $kv_A/\Omega\approx
H/(\lambda\beta^{1/2})$.  Long-wavelength modes for which $\lambda/H
\gg \beta^{-1/2}$ are convective, while modes for which $\lambda/H
\sim \beta^{-1/2}$ are magnetorotational.  The convective modes are
primarily driven by buoyancy, not magnetic tension (Fig. 2).
Moreover, they transport angular momentum {\it inwards} rather than
outwards (Fig. 1b), just as indicated by hydrodynamic studies.  For
the long-wavelength modes to be of interest the accretion flow must
have $\beta \gg 1$ so that the convective fluctuations fit inside the
disk.  Under these conditions we believe that the hydrodynamic CDAF
models can accurately describe the structure of the accretion flow.
The precise $\beta$ that allows a hydrodynamical analysis is uncertain
and can be determined only by magnetohydrodynamic simulations (the
linear analysis suggests that $\beta \gsim 10$ may suffice).

It is interesting to note that the saturation $\beta$ of the MRI does
not appear to be universal.  Given the plausible dependence on $\beta$
highlighted above, this may have important implications for the
structure of radiatively inefficient accretion flows.
Three-dimensional MHD simulations find that that if the field is
initially toroidal a radiatively inefficient accretion flow develops
into a CDAF-like state (Igumenshchev, Abramowicz \& Narayan 2002),
while if the field has a vertical component the structure is very
different (Hawley, Balbus \& Stone 2001; Hawley \& Balbus 2002; see
also Stone \& Pringle's 2001 2D simulations).

To conclude, it is important to emphasize that the primary conclusion
of the analytic arguments in this paper is not that CDAFs {\it
actually do} describe the structure of radiatively inefficient
accretion flows, but rather that they {\it could}.  In particular, our
linear MHD analysis supports all of the key CDAF assumptions so long
as $\beta \gg 1$.

\acknowledgements

RN, IVI, and EQ were supported in part by NSF grant AST-9820686, RFBR
grant 00-02-16135, and NASA grant NAG 5-12043, respectively.


\clearpage
\begin{appendix}
\section{\bf Appendix}
\bigskip
In the main text we focused exclusively on unstable modes and
therefore considered only the positive sign of the root in equation
(\ref{growth}).  Here we generalize the discussion and consider both
unstable and stable modes.  To this end, we rewrite equation
(\ref{growth}) as follows \beq \gamma^2 = -(k v_A)^2 +
{(N^2-\kappa^2)\over2}\left[1 \pm \sqrt{1 +
16\Omega^2(kv_A)^2/(N^2-\kappa^2)^2}\right].\label{disppm} \eeq The +
and $-$ signs in equation (\ref{disppm}) map directly onto the + and
$-$ signs in equation (\ref{growth}) when $N^2-\kappa^2>0$, but the
mapping is reversed when $N^2-\kappa^2<0$.  Equation (\ref{disppm})
is, of course, completely equivalent to equation (\ref{growth}), but
for this Appendix we find it more convenient to work with the $\pm$
convention in equation (\ref{disppm}) because of its behavior when
$N^2 - \kappa^2$ changes sign.  In what follows, we refer to the +
sign in equation (\ref{disppm}) as the ``+ mode'' and the $-$ sign as
the ``$-$ mode.''

Let us focus on the long-wavelength limit $kv_A/\Omega \ll 1$.  The
growth rates of the + and $-$ modes are then given by \beq \gamma_+^2
\approx (N^2-\kappa^2) + \left( {4\Omega^2\over N^2-\kappa^2} -1
\right )(kv_A)^2, \label{plus} \eeq \beq \gamma_-^2 \approx - \left(
{4\Omega^2\over N^2-\kappa^2} + 1 \right )(kv_A)^2.
\label{minus}
\eeq 

In the limit $kv_A/\Omega \ll 1$, the growth rate of the + mode is the
same as in hydrodynamics: $\gamma_+^2=N^2-\kappa^2$.  If
$N^2>\kappa^2$ (H{\o}iland unstable), $\gamma_+^2$ is positive and the
mode is unstable, while if $N^2<\kappa^2$, $\gamma_+^2$ is negative
and the mode is oscillatory in nature.  In either case, the mode is
completely insensitive to the magnetic field.

The $-$ mode, on the other hand, has a growth rate that vanishes as
$kv_A\to0$, which is a clear indication that this mode is strongly
influenced by the magnetic field.  Indeed, the mode owes its very
existence to the presence of the magnetic field.  The $-$ mode is
associated with the classic MRI of BH91.

An inspection of equations (\ref{plus}) and (\ref{minus}) reveals the
following interesting result: assuming that $|N^2-\kappa^2| <
4\Omega^2$ (which is almost always the case), only one of the two
modes can be unstable.  When the medium is convectively stable ($N^2 <
\kappa^2$), the + mode (which we argued above is hydrodynamic in
nature) is obviously stable.  In this case, the $-$ mode is unstable
and gives the long-wavelength limit of the MRI.  However, when the
medium is convectively unstable ($N^2 > \kappa^2$), then it is the +
mode that grows, while the $-$ mode is stable.  Thus, in the long
wavelength limit, the MRI actually disappears if the medium is
unstable by the H{\o}iland criterion.  Fortunately, this is precisely
when the medium is convectively unstable so there are always unstable
modes available, regardless of the sign of $N^2-\kappa^2$.

\begin{figure}
\plotone{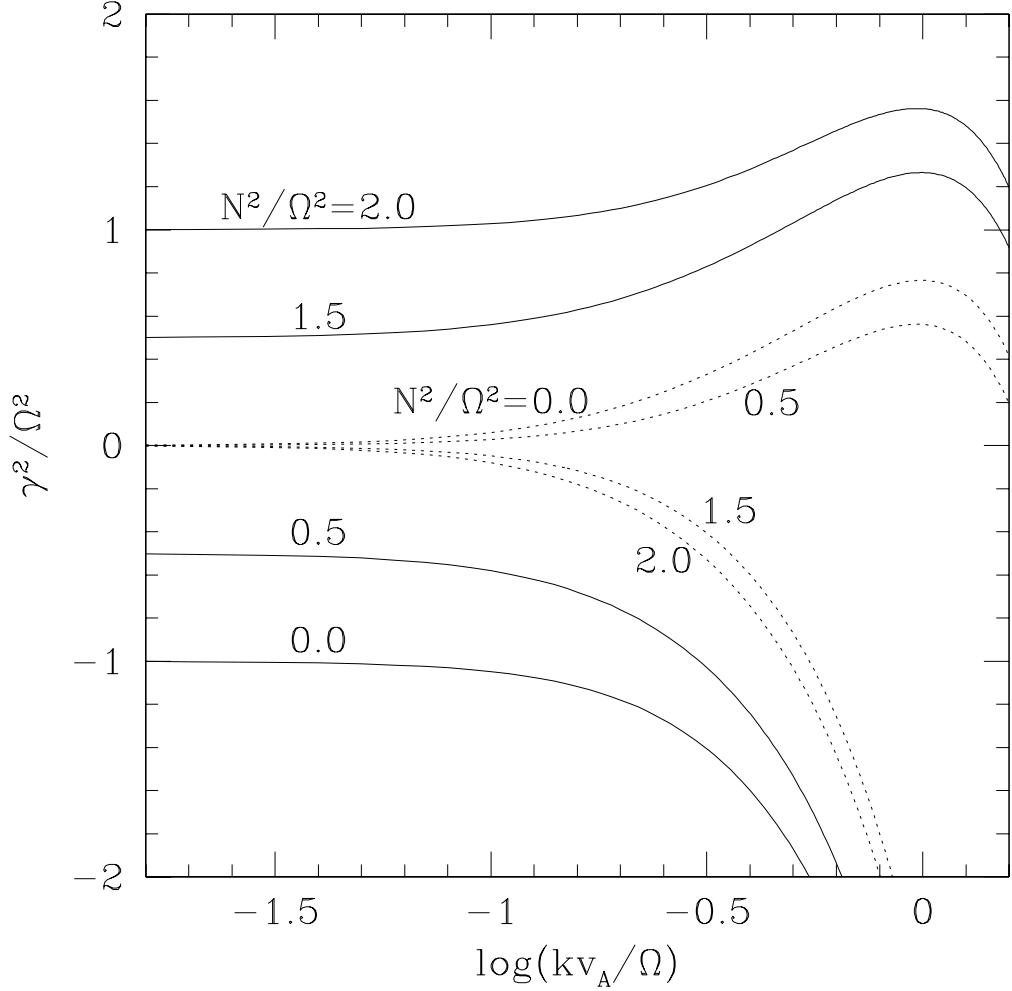}
\caption{The dimensionless growth rate $\gamma^2/\Omega^2$ versus
dimensionless wavevector $kv_A/\Omega$ for $N^2/\Omega^2=0, ~0.5,
~1.5, ~2$.  A mode with a positive value of $\gamma^2$ is unstable,
while one with a negative value of $\gamma^2$ is stable (oscillatory).
The solid and dashed lines correspond to the + and $-$ modes,
respectively, in equation (\ref{disppm}).  Each curve is labeled by
the value of $N^2/\Omega^2$.
\label{fig3}}
\end{figure}

\placefigure{fig3}

Figure (\ref{fig3}) shows $\gamma^2$ as a function of $(kv_A)^2$ for
the + and $-$ modes for representative values of $N^2/\Omega^2$.  We
see that modes that are stable ($\gamma^2<0$) in the limit $kv_A\to0$
remain stable for all values of $k$, while modes that are unstable at
small values of $kv_A$ reach a maximum growth rate for $kv_A \sim
\Omega$ and then become stable with increasing $kv_A$.

\end{appendix}

\clearpage

\begin{thebibliography}{999}

\bibitem[]{} Abramowicz, M. A., Chen, X.-M., Kato, S., Lasota, J.-P., 
\& Regev, O. 1995, \apj, 438, L37
\bibitem{} Abramowicz, M. A., Igumenshchev, I. V., Quataert, E., \& Narayan, 
R. 2002, ApJ, 565, 1101
\bibitem[Balbus(1995)]{1995ApJ...453..380B} Balbus, S.~A.\ 1995, \apj, 453, 380
\bibitem[Balbus(2000)]{2000ApJ...534..420B} Balbus, S.~A.\ 2000, \apj, 534, 420
\bibitem[Balbus(2001)]{2001ApJ...562..909B} Balbus, S.~A.\ 2001, \apj, 562, 909
\bibitem[Balbus \& Hawley(1991)]{1991ApJ...376..214B} Balbus, S.~A.,~\& 
Hawley, J.~F.\ 1991, \apj, 376, 214 (BH91)
\bibitem[Balbus \& Hawley(1992)]{1992ApJ...392..662B} Balbus, S.~A.,~\& 
Hawley, J.~F.\ 1992, \apj, 392, 662
\bibitem[Balbus \& Hawley (1998)]{bh97} Balbus, S.A., \& Hawley, J.F. 1998, 
Rev. Mod. Phys., 70, 1
\bibitem{} Balbus, S. A., \& Hawley, J. F., 2002, \apj, submitted, 
astro-ph/0201428 (BH02)
\bibitem[]{} Dorland, W., Jenko, F., Kotschenreuther, M., \& Rogers, B. N.
2000, Phys. Rev. Lett., 85, 5579
\bibitem{} Goldreich, P., \& Shubert, G., 1967, ApJ, 150, 571
\bibitem{} Hawley, J. F. \& Balbus, S. A., 2002, ApJ in press (astro-ph/0203309)
\bibitem[Hawley, Balbus, \& Stone(2001)]{2001ApJ...554L..49H} Hawley, J.~F., 
Balbus, S.~A., \& Stone, J.~M.\ 2001, \apjl, 554, L49
\bibitem[]{} Ichimaru, S. 1977, \apj, 214, 840
\bibitem[]{} Igumenshchev, I. V., \& Abramowicz, M. A. 1999, \mnras, 303, 309
\bibitem[]{} Igumenshchev, I. V., \& Abramowicz, M. A. 2000, \apjs, 130, 463
\bibitem[]{} Igumenshchev, I. V., Chen, X., \& Abramowicz, M. A. 1996, \mnras, 
278, 236
\bibitem[]{} Igumenshchev, I. V., Abramowicz, M. A., \& Narayan, R. 2000,
\apj, 537, L27
\bibitem[]{} Igumenshchev, I. V., Abramowicz, M. A., \& Narayan, R. 2002,
in preparation
\bibitem[]{} Narayan, R. 2002, in {\it Lighthouses of the Universe},
ed. M. Gilfanov, \& R. Sunyaev (Heidelberg: Springer-Verlag), in press,
astro-ph/0201260
\bibitem[]{} Narayan, R., Igumenshchev, I. V., \& Abramowicz, M. A.
     2000, \apj, 539, 798
\bibitem[]{} Narayan, R., Mahadevan, R., \& Quataert, E. 1998, in {\it
Theory of Black Hole Accretion Discs}, ed. M. A. Abramowicz,
G. Bjornsson, \& J. E. Pringle (Cambridge: Cambridge Univ. Press), p148
\bibitem[]{} Narayan, R., \& Yi, I. 1994, \apj, 428, L13
\bibitem[]{} Quataert, E. 2001, in {\it Probing the Physics of AGN
with Multiwavelength Monitoring}, ed. B. M. Peterson, R. S. Polidan, \&
R. W. Pogge (San Francisco: Astr. Soc. Pacific), p71q
\bibitem[]{} Quataert, E., \& Gruzinov, A. 2000, \apj, 539, 809
\bibitem[]{} Rogers, B. N., Dorland, W., \& Kotschenreuther, M. 2000,
Phys. Rev. Lett., 85, 5536
\bibitem[Ryu \& Goodman(1992)]{1992ApJ...388..438R} Ryu, D.,~\& Goodman, 
J.\ 1992, \apj, 388, 438
\bibitem[]{} Stone, J. M., Pringle, J. E., \& Begelman, M. C. 1999, \mnras, 
310, 1002
\bibitem{} Stone, J. \& Pringle, J. E., 2001, MNRAS, 322, 461
\bibitem{} Tassoul, J.-L., 1978, {\it Theory of Rotating Stars} 
(Princeton: Princeton University Press)

\end{thebibliography}
\end{document}